\documentclass[floatfix,aps,prc,noeprint,twocolumn,table,nofootinbib]{revtex4-1} % two column journal format % two column journal format
\usepackage{amsfonts}                            % Use AMS fonts
\usepackage{amsmath}
\usepackage{amssymb}
\usepackage{textcomp}
\usepackage[compat=1.0.0]{tikz-feynman}
\usepackage{mathtools}                           % and definitions
\usepackage{bm}
\usepackage{graphicx}                            % Graphics package
\usepackage{color}
\usepackage{latexsym}                            % Load additional symbols
\usepackage{dcolumn}                             % Decimal-dot aligned columns
\usepackage[colorlinks=true,linkcolor=blue,citecolor=blue,urlcolor=blue]{hyperref}
\usepackage[normalem]{ulem}
\usepackage{slashed}
\usepackage{xfrac}
\usepackage{cleveref}
\crefname{figure}{Fig.}{Figs.}
\crefname{equation}{Eq.}{Eqs.}
\usepackage{natbib}
\usepackage{mathrsfs}
\usepackage{physics} 
\usepackage{comment}
\usepackage[caption=false]{subfig}
\usepackage{dynkin-diagrams}
\def\row#1/#2!{#1_{\IfStrEq{#2}{}{n}{#2}} & \dynkin{#1}{#2}\\}
\usepackage[all,cmtip]{xy}
\usepackage{tikz-cd}
\usepackage{tikz}

\begin{document}

\title{A Convenient Representation Theory of Lorentzian Pseudo-Tensors:\newline $\mathcal{P}$ and $\mathcal{T}$ in $\operatorname{O}(1,3)$}
%\author{Author}
\author{Craig M$^{\mathrm{c}}$Rae}
%\affiliation{Department of Physics and Astronomy}
\affiliation{\mbox{Department of Physics and Astronomy,
	University of Manitoba}, Winnipeg, Manitoba, Canada R3T 2N2}

\begin{abstract}
A novel approach to the finite dimensional representation theory of the entire Lorentz group $\operatorname{O}(1,3)$ is presented. It is shown how the entire Lorentz group may be understood as a semi-direct product between its identity component and the Klein four group of spacetime reflections: $\operatorname{O}(1,3) = \operatorname{SO}^+(1,3) \rtimes \operatorname{K}_4$. This gives way to a convenient classification of tensors transforming under $\operatorname{O}(1,3)$, namely that there are four representations of $\operatorname{O}(1,3)$ for each representation of $\operatorname{SO}^+(1,3)$, and it is shown how the representation theory of the Klein group $\operatorname{K}_4$ allows for simple book keeping of the spacetime reflection properties of general Lorentzian tensors, and combinations thereof, with several examples given. There is a brief discussion of the time reversal of the electromagnetic field, concluding in agreement with standard texts such as Jackson \cite{JACKSON3RD}, and works by Malament \cite{MALAMENT2004}. 
\end{abstract}

%SigStatement: Students and experts alike are often confused about the interaction of the discrete spacetime symmetries P and T with the more well understood Lorentz invariance of boosts and rotations. For example there is still work being published debating how the electromagnetic fields should transform under time reversal. This is due in part to a lack of a convenient formalism relating the two, leaving all calculations of this type to be done in an ad hoc or case-by-case basis. This article provides such a formalism for an overarching framework to consistently and conveniently discuss and keep track of the the properties of various physical quantities under space-time reflections.
\maketitle
\tableofcontents

\newpage

\section{\texorpdfstring{What do we mean by `Time Reversal'?}{What do we mean by 'Time Reversal?'}}
\vspace{-0.8em}
Time reversal is one of the most confused topics in modern physics, with students and experts alike often having little understanding of the intricacies involved in defining the notion, so it is worth specifying precisely what is meant by it. Most sources will simply point to the progenitor of the modern understanding of time reversal, Wigner's exploration of the topic in the 1930's \cite{Wigner1932}, \cite{WIGNERBOOK}. While certain notions have been set in stone since (such as time reversal's necessary anti-unitarity), there remains a great deal of confusion about particular interpretations and applications of the concept. Despite physics textbooks being largely in agreement about the effects a `time reversal operator' should have on any particular physical phenomenon, it is often treated case-by-case, and there remains no universally agreed upon definition for what physicists and philosophers of physics mean when they discuss the `time reversal' of a system. This is no better highlighted than in the disagreements and debate between, for example, Malament \cite{MALAMENT2004}, Arntzenius \& Greaves \cite{GREAVESARNTZENIUS}, Roberts \cite{ROBERTS2022}, Albert \cite{ALBERT2003}, and Callender \cite{Callender}. It is this confusion which led me to attempt to understand the discrete spacetime symmetries in a self consistent way. I believe the attempt to have been successful, making precise many details lacking in textbook explanations, and delivering a convenient formalism for keeping track of the properties of physical quantities under spacetime reflections. 

What is meant mathematically in this paper by `time reversal' will be made clear in Sec.~(\ref{sec:RepsofPandT}). In words however, it should be understood literally as a reflection directly analogous to parity. To give a simple example, suppose we have an ordering of objects in space, three spheres identical but in color named $r,g,b$, placed at coordinates $(1,2,3)$ along an axis directly in front of us. Placing a mirror normal to the axis at coordinate $(4)$ produces an image of the spheres under the operation of parity, the images appearing at coordinates $(5,6,7)$ with colors $b,g,r$ respectively. When we talk about a system, object, theory, etc. being parity symmetric, we mean that were we to understand the parity reflected system as if it were a system within our world, the two systems appear identical, or essentially identical up to the application of other agreed upon transforms and symmetries. In the example given the parity reflected state is not strictly symmetric, unless we mean symmetric up to translations and rotations, since the order of the colors has been reversed, and the images coordinates are different than the originals. Were we to replace the third sphere $b$ with another sphere $r$, then the state is manifestly symmetric (or invariant) under a reflection through coordinate $2$: i.e. the parity reflection of this setup, understood as another possible setup of the objects in our world, is identical to the initial setup of the objects. Importantly, note that when we consider the parity reflected state, we do not parity transform ourselves as well. 

The philosophy is the same when it comes to temporal reflection. The one dimensional nature of time necessitates a partial ordering of events for any observer, and so under a temporal reflection, what is to be inspected for symmetry or invariance, is the set of events in reversed order, but understood as a possible sequence of events flowing forward in time. To be explicit here let it be said we \textit{do not} interpret a time reversal of an ordered set of events, as the events evolving backward in time. We interpret it rather as a possible forward time evolution of some (possibly differently prepared) system. 

Finally in the work below, what is meant by `time reversal invariance', `parity symmetry', and synonyms thereof, are not the strict examples provided above: it is not that states or systems are essentially identical before and after the application of the reflection. What is more interesting physically and typically what is meant by these terms, is that given a physical theory describing the evolution of a system: a theory exhibits parity symmetry if for every solution to the theory's equations of motion, any parity reflection of the solution is also a solution to the equations of motion. Directly analogously a theory is said to exhibit \textit{time reversal invariance} if for every solution to the theory's equations of motion, any temporally reflected solution is a solution to the equations of motion. 

In less precise language, for a system described by some physical theory, if all predicted behaviors of the system as seen in a (any) mirror are also possible behaviors of the system, then the theory is parity symmetric. Likewise if for all predicted behaviors of the system, when the order of events is reversed these temporally reflected behaviors are also possible behaviors of the system, the theory is time reversal symmetric.\footnote{The author presently only has classical and semi-classical pictures in mind. The full quantum mechanical case requires additional care and detail, for several reasons, foremost among them being that the discussion becomes about operators, state spaces, and observables, instead of trajectories and values of fields.} Maxwell's equations, for example, exhibit both of these symmetries; a proof of this in given in appendix.~(\ref{subsubsec:Maxwell}).

%\subsection{Passive Reflections}
%What is usually meant by parity symmetry and time reversal invariance, is an active transformation, wherein one imagines a universe which has been reorganized by the reflection. The much less interesting but mathematically equivalent formulation is the passive case, where we simply inspect the changes of components and coordinates of objects under the formal redefinition of coordinates from $x^{0} \mapsto -x^0$ or $x^{i} \mapsto -x^{i}$. Of course these choices are completely arbitrary, and so in this case the breaking of these symmetries would have to be understood as the relevant physical equations not coming back to themselves under this redefinition. See Sec.~(\ref{heateq}) for an example of this kind of failure.
%\vspace{-0.8em}

\vspace{-1em}
\subsection{\label{subsec:Velocities}Simplest Motivating Example: Velocities}
\vspace{-0.6em}
\begin{figure}[!b]
\centering
\includegraphics[width=0.45\textwidth]{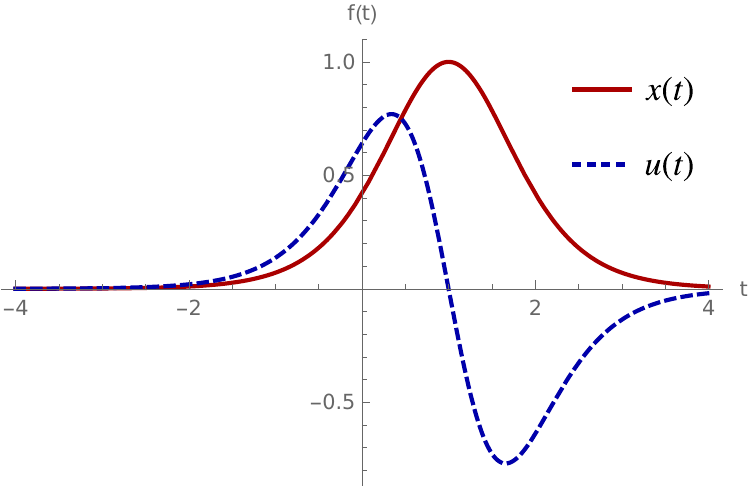}\hfill
\includegraphics[width=0.45\textwidth]{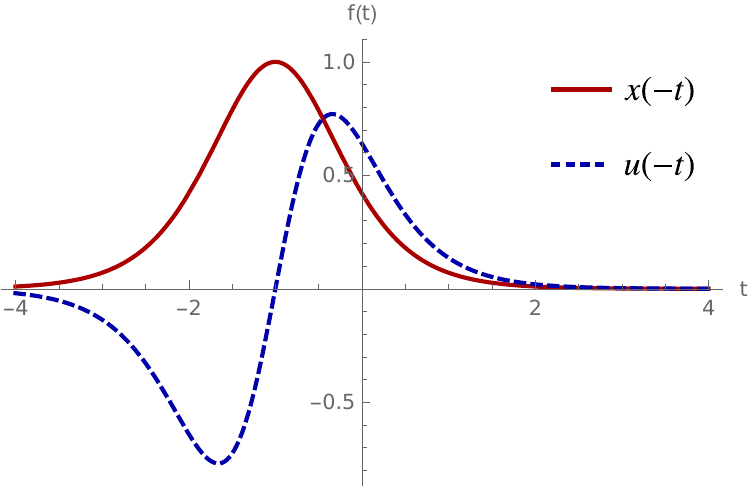}\hfill
\includegraphics[width=0.45\textwidth]{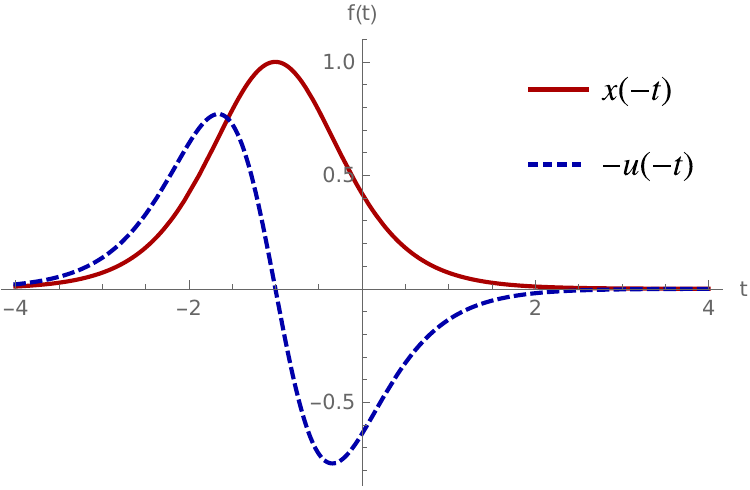}
\caption{In solid red we have a trajectory $x(t) = \operatorname{sech}^2(t-1)$. In the top panel we see the trajectory plotted with its velocity (dashed blue). In the middle panel we see the naive reflection of both functions. This naive reflection does not send the velocity function to the velocity function of the reflected trajectory. Finally on the bottom we see the negation ensures the velocity is mapped correctly under reflection.}
\label{fig:simpletraj}
\end{figure}
Before we come to the formalities of group and representation theory in the following sections, it is illustrative to see how something as simple as velocities lead naturally to a consideration of something like `parity' or `charges' under reflections of coordinates. Given a one dimensional trajectory $x(t)$, and its velocity function $u(t) = dx/dt$, a temporal reflection about $t=0$ gives us the corresponding function $x(-t)$. If we were naively given the function $u(t)$ and asked to perform a time reflection of it, we would just as well return the function $u(-t)$. However, this function is not the velocity of the time reversed trajectory: 
\begin{equation}
\frac{dx(-t)}{dt} = -u(-t),
\end{equation}
its negation is, as illustrated in Fig.~(\ref{fig:simpletraj}).

Of course we can see the chain rule from the time derivative is what causes the additional sign flip, but often we find ourselves in situations where this `implicit' additional time dependence is not obvious. Should we wish to understand the time reversal of some complicated dynamics, it is cumbersome to unpack the definition of each object, to see if a possible chain rule might apply. Integrations and and derived quantities thereof generally should invite us to anticipate the possibility of objects which pick up a sign under temporal reflections whether or not their definition explicitly contains a time coordinate or time derivative. This motivates the notion that some objects are `charged' or `odd' under time reversal, and some objects are `uncharged' or `even'; no different from how the parity of a vector distinguishes coordinate vectors and pseudo-vectors. We will see that treating this notion rigorously allows for simple book keeping of quantities which physically ought to pick up additional signs under spacetime reflections. In this language, under time reversal positions would be uncharged (even) and velocities would be charged (odd), whereas they are both charged (odd) under spacial inversion. 

\vspace{-1em}
\section{\texorpdfstring{The Structure of O(1,3)}{The Structure of the Entire Lorentz Group}}
\vspace{-0.6em}
\label{sec:RepsofPandT}
\begin{figure}[h]
\centering{
\begin{tikzpicture}
%Identity
\node[draw,
	circle,
	minimum size=0.9cm,
	fill=white
] (id) at (0,0){$1$};
\node[] at (id.center){$1$};
% 3-reflection
\node[draw,
	circle,
	minimum size=0.9cm,
	fill=white,
    right=2cm of id
] (eta) {$\mathcal{R}$};
%1-reflection
\node[draw,
	circle,
	minimum size=0.9cm,
	fill=white,
    below=2cm of id
] (minuseta) {$-\mathcal{R}$};
%minusid
\node[draw,
	circle,
	minimum size=0.9cm,
	fill=white,
    right=2cm of minuseta
] (minusid) {$-1$};
% Arrows with text label
\draw[-stealth] (id.east) -- (eta.west)
	node[midway,above]{$\mathcal{R}$};
\draw[-stealth] (eta.west) -- (id.east)
    ;
\draw[-stealth] (id.south) -- (minuseta.north)
	node[midway,left]{$-\mathcal{R}$};
\draw[-stealth] (minuseta.north) -- (id.south)
    ;
\draw[-stealth] (minusid.north) -- (eta.south)
	node[midway,right]{$-\mathcal{R}$};
\draw[-stealth] (eta.south) -- (minusid.north)
    ;
\draw[-stealth] (minuseta.east) -- (minusid.west)
	node[midway,below]{$\mathcal{R}$};
\draw[-stealth] (minusid.west) -- (minuseta.east)
    ;
\draw[-stealth] (id.south east) -- (minusid.north west)
	node[midway,below]{};
\draw[-stealth] (minusid.north west) -- (id.south east)
    ;
\draw[-stealth] (eta.south west) -- (minuseta.north east)
	node[midway,right]{$-1$};
\draw[-stealth] (minuseta.north east) -- (eta.south west)
    ;
\end{tikzpicture}
}
\caption{The four connected components of the group $\operatorname{O}(1,3)$, along with the reflections mapping between these components: $\mathcal{R}$ = $\operatorname{diag}\{1,-1,-1,-1\}$.}\vspace{-0.6em}
\label{LorentzComp}
\end{figure}

In the following sections a convenient representation theory of Lorentzian pseudo-tensors\footnote{A pseudo-tensor is used here in the sense as pseudo-vector and pseudo-scalar: additional sign changes under the action of reflections. Axial and pseudo are synonymous prefixes.} is constructed: a way of labeling physical quantities so their the parity and time reversal properties are transparent to us at a glance. For brevity it is assumed the reader is familiar with special relativity, in particular the ideas of inertial observers, a lack of preferred reference frames, the notion of Lorentz invariance, and boosts and rotations as the building blocks of Lorentz transforms. 

This approach comes out of the representation theory of the entire Lorentz group: $\operatorname{O}(1,3)$. The word \textit{entire} here is used to distinguish from the usual abuse of language whereby physicists say `The Lorentz Group' to mean only the component of the entire Lorentz group connected to the identity transform. 
\begin{equation}
\operatorname{SO}^{+}(1,3) \equiv \operatorname{PSO}(1,3).
\end{equation}
The identity component is known to its friends as the proper orthochronus Lorentz group, a mathematician would call it the projective special orthogonal group. The entire group has four connected components, disjoint from one another, illustrated in Fig.~(\ref{LorentzComp}). Note that the usual identifications of the parity operator $\mathcal{P}$ with $\mathcal{R}$ and the time reversal operator $\mathcal{T}$ with $-\mathcal{R}$, are avoided. The reason for this avoidance is precisely because the representation theory of the entire Lorentz group is more subtle than this simple assignment of $\mathcal{P}$ and $\mathcal{T}$; we will come to see these operators behave differently when acting upon distinct representations, just as say, familiar angular momenta operators look different acting upon different representations. 

Any Lorentz transform in $\operatorname{SO}^+(1,3)$, the identity component, can be reached by appropriate choice of angle and boost parameter.\footnote{Formally this is declaring the fact that the map from the Lie algebra to the identity component of the Lie group is surjective. It is true here but not always the case, such as for $\operatorname{SL}(2,\mathbb{C})$} These transforms preserve orientations in space and the sense of time. Elements within the other three disconnected components may then be reached by multiplication of one of the reflections: $\{\mathcal{R}, -\mathcal{R},-1\}$, which allow for changes of frame which swap the handedness of a coordinate system, or reverse the sense of time. These three reflections together with the identity form a group known as the Klein group $\operatorname{K}_4$, discussed next in Sec.~(\ref{subsec:Klein}). 

The identity component is the piece of the entire group for which the standard representation theory is well understood. What we are missing then is a compelling representation theory of the three other components of the entire Lorentz group; but these components are merely `reflected copies' of the identity component. Therefore what is desired is a way to think of the entire Lorentz group as some product of the piece we understand, and some discrete reflection group which generates these reflected copies. In what is to follow we will see the entire Lorentz group may be understood as the semi-direct product of groups\footnote{See Appendix~(\ref{SemiDirect}) for a review of semi-direct products.}
\begin{equation}
\operatorname{O}(1,3) \cong \operatorname{SO}^{+}(1,3) \rtimes \operatorname{K}_4.
\end{equation}
\noindent
Again $\operatorname{K}_4$ is the Klein four group \cite{vier}, corresponding to the action of the discrete reflection operators $\mathcal{P},\mathcal{T}$, and $\mathcal{PT}$, explored in detail in the next section. We will often refer to this group as ``the discrete reflection group". What is particularly important about the above decomposition of the entire Lorentz group, is that the representations of a products of groups can often be given by the tensor product of representations of said groups, and this is true here. Therefore to understand representations of the entire Lorentz group, we must understand the representations of $\operatorname{K}_4$. The physical distinction between say, a pseudo-four-vector and a coordinate four-vector, can only be captured by understanding the representation theory of the entire Lorentz group, and failing to do this can lead to misunderstandings about the nature of objects under the operation of $\mathcal{P}$ and $\mathcal{T}$.

\vspace{-0.8em}
\subsection{\label{subsec:Klein}The Klein Four Group}
\vspace{-0.6em}
The Klein four group $\operatorname{K}_4$, also called the Vierergruppe, abstractly is a finite Abelian group containing four elements, suggestively named $\{1,\mathcal{P},\mathcal{T},\mathcal{PT}\}$ satisfying:
\begin{equation}
\label{K4def}
\mathcal{P}^2 = \mathcal{T}^2 = (\mathcal{PT})^2 = 1, \quad \mathcal{P}\cdot \mathcal{T} = \mathcal{PT}.
\end{equation}
As $\operatorname{K}_4$ is Abelian, Schur's lemma \cite{WuKiTung} tells us its irreducible representations over $\mathbb{C}$ are all one dimensional. For $\operatorname{K}_4$ this remains true even over $\mathbb{R}$. There are exactly four distinct irreducible representations of the Klein group, given in Tab.~(\ref{K4reps}).

\begin{table}[h!]
\centering {\large
\begin{tabular}{ c|c|c|c|c } 
 & $1$ & $\mathcal{P}$ & $\mathcal{T}$ & $\mathcal{PT}$\\ 
 \hline
 $\rho_1$ & $\>\>\>1$ & $\>\>\>1$ & $\>\>\>1$ & $\>\>\>1$\\ 
 \hline
 $\rho_{T}$ & $\>\>\>1$ & $\>\>\>1$ & $-1$ & $-1$ \\ 
 \hline
 $\rho_P$ & $\>\>\>1$ & $-1$ & $\>\>\>1$ & $-1$ \\ 
 \hline
 $\rho_{PT}$ & $\>\>\>1$ & $-1$ & $-1$ & $\>\>\>1$ \\ 
 \hline
\end{tabular}}
\caption{The four irreducible real representations of $\operatorname{K}_4$, named here: $\rho_1$, $\rho_T$, $\rho_P$, and $\rho_{PT}$, given by the explicit representation maps $\rho: \operatorname{K}_4 \mapsto \mathbb{R}^*$.}
\label{K4reps}
\end{table}

The representations here have been so named for two reasons. Firstly if an element transforms under the action $\rho_T$, this means it is `charged' under operators containing $\mathcal{T}$, and invariant otherwise; analogously for $\rho_P$. Objects which transform under the representation $\mathcal{\rho}_{PT}$ are to be understood as being charged under both $\mathcal{P}$ and $\mathcal{T}$, and so $\mathcal{P}\mathcal{T}$ acts as the identity. I.e. we may uniquely identify the representations by the which of $\mathcal{P}$ and $\mathcal{T}$ act as $-1$ in that representation. 

The second reason for the naming, is the wonderful fact that tensor products of representations of $\operatorname{K}_4$, form the group $\operatorname{K}_4$ themselves!\footnote{The cause of this property is due to the fact finite Abelian groups are self dual, under Pontryagin duality, i.e. the character group of a finite Abelian group is (non-canonically) isomorphic to the group \cite{Pontry}. } If we have an object $A$ which transforms under $\rho_P$ and an object $C$ which transforms under $\rho_{PT}$, then the tensor $A \otimes C$ must transform under $\rho_T$, etc. Given the four representations and the defining relations of $\operatorname{K}_4$, we have precisely everything we need in order to understand not only the abstract group, but also the representation theory of our discrete reflection group.

For convenience let us introduce a charge operator $K$ for the Klein group: when given an object, $K$ returns the representation of the Klein group the object transforms under, i.e. its `charges' or parities under the discrete reflection group. In the following section we will consider larger representations of $\operatorname{K}_4$; these must take the form of direct sums of one dimensional representations. From here on, we will refer to representations simply by their odd charges, i.e. their subscript $1, P, T, PT$, and the abstract operators (the things being represented) will be written $\mathcal{P}$ and $\mathcal{T}$. 
\vspace{-1em}
\subsection{\label{subsec:Def}The Defining Representation of \texorpdfstring{$\operatorname{O}(1,3)$}{O(1,3)}}
\vspace{-0.6em}
The defining representation of $\operatorname{O}(1,3)$ acting on $\mathbb{R}^{1,3}$ with Cartesian coordinates $x^\alpha = (t, \mathbf{x})$ identifies the parity operator $\mathcal{P}$, and the time reversal operator $\mathcal{T}$ acting on these coordinates via the matrices $\mathcal{P}_{\operatorname{def}}$ and $\mathcal{T}_{\operatorname{def}}$:
\begin{equation}
\mathcal{P}_{\mathrm{def}} = 
\begin{pmatrix}
1 & 0 & 0 & 0 \\
0 & -1 & 0 & 0 \\
0 & 0 & -1 & 0 \\
0 & 0 & 0 & -1 
\end{pmatrix}, \hfill\>\> \mathcal{T}_{\mathrm{def}} = 
\begin{pmatrix}
-1 & 0 & 0 & 0 \\
0 & 1 & 0 & 0 \\
0 & 0 & 1 & 0 \\
0 & 0 & 0 & 1 
\end{pmatrix}.
\end{equation}
This representation is given by one component which is odd under time reversal (the time coordinate), and three components which are odd under parity (the spacial coordinates). The reflection charges of the vector are:
\begin{equation}
K[x^\alpha] = T\oplus P\oplus P\oplus P = \begin{pmatrix} T\\
P\\
P\\
P
\end{pmatrix} := 
\begin{pmatrix} T\\
\vec{P}
\end{pmatrix} .
\end{equation}
Let us inspect something more interesting: the momentum of a point particle. Suppose a point particle has a trajectory given in coordinates by
\begin{equation}
X^\alpha(t) = \begin{pmatrix}
t \\
\mathbf{X}(t)
\end{pmatrix}, \quad K[X^\alpha] = \begin{pmatrix} T\\
\vec{P}
\end{pmatrix}.
\end{equation}
The four-momentum is given by a mass $\mu$ multiplied into the proper time derivative of the trajectory. By assumption the mass is invariant under all reflections in $\operatorname{K}_4$, and the time derivative is necessarily odd under time reversal:
\begin{equation}
K[\mu] = 1, \quad  \mathrm{and} \quad K\left[\frac{d}{d\tau}\right] = T.
\end{equation}
With this we find that our four-momentum
\begin{equation}
p^\alpha = \mu \frac{dX^\alpha}{d\tau} = \begin{pmatrix} E \\ 
\mathbf{p}\end{pmatrix},
\end{equation}
is charged under the reflection group as
\begin{equation}
\begin{split}
K[p^\alpha] &= K[\mu] \otimes K\left[\frac{d}{d\tau}\right] \otimes K[X^\alpha] \\
&= 1 \cdot T \cdot  \begin{pmatrix} T \\ 
\vec{P}\end{pmatrix} = \begin{pmatrix} T^2 \\ 
T\cdot\vec{P}\end{pmatrix} = \begin{pmatrix} 1 \\ 
\vec{PT}\end{pmatrix} .
\end{split}
\end{equation}

This tells us that the energy is completely invariant under the action of the discrete reflection group $\operatorname{K}_4$, and the three-momenta are odd under both parity and time reflections. This matches our intuition, and is the first example of a failure of the `standard' defining assignments of the parity and time reversal operators. In this representation both $\mathcal{P}$ and $\mathcal{T}$ are represented by the same matrix, `$\mathcal{R}$' (defined above), and $\mathcal{PT}$ acts as the identity. 

Next we need to make sense of how the discrete reflection group $\operatorname{K}_4$ interacts with the proper orthochronus Lorentz group $\operatorname{SO}^+(1,3)$, especially under the action of boosts which at present would seem to `mix up' our clean one dimensional decomposition of these four-vectors, perhaps leading to different observers ascribing different charges under the reflection group to the same physical objects and causing a contradiction. The next two sections address this concern, constructing precisely those combinations of reflection charges which have physically reasonable transformation properties.
\vspace{-0.6em}
\subsection{\label{subsec:Boosts}The Effects of Parity and Time Reversal on Boosts and Rotations}
\vspace{-0.6em}
The interaction of boosts and rotations with the basic spacetime reflections is mostly straightforward. Suppose in one frame we have a particle traveling along the $x$ axis with velocity $\beta = v/c$; we may transform our coordinates to a co-moving frame by application of the boost $B(\beta) = e^{\omega K_x}$, where $\omega = \operatorname{arctanh}(\beta)$ is the rapidity, and $K_x$ is the relevant generator of boosts defined in Appendix~(\ref{sec:AutLor}). Under either a parity inversion or time reversal of this setup, the velocity of the reflected system will be reversed compared to the original, and so the correct boost to enter the co-moving frame after either reflection will be given via $B(-\beta) = e^{-\omega K_x}$. Of course, if one boosts first, and then reflects, it is the same as reflecting first and then boosting in the reversed direction. This can be phrased via commutation relations: 
\begin{equation}
\mathcal{P} B(\beta) = B(-\beta)\mathcal{P}, \quad \mathcal{T} B(\beta) = B(-\beta)\mathcal{T}.
\end{equation}

That is to say under a spacial or temporal reflection the boost parameter will pick up a sign, just as a velocity will. In the language introduced above, we would say boost parameters (rapidities, velocities etc.) are charged under the $PT$ representation of the reflection group. With an understanding of how $\mathcal{P}$ and $\mathcal{T}$ interact with boosts, next we turn to rotations. It turns out rotations commute with parity and time reversal. As a generic example, suppose one hopes to make some vector co-linear with a specified target direction, via a rotation. In 3D any rotation can be understood as a rotation in a single plane, i.e. around a single axis; the plane in this case is the span of the original vector and the target direction. If the vector is odd under $\mathcal{T}$, the negation of the vector and the target vector is equivalent to a rotation by $\pi$ in that plane, and all rotations in the plane commute. If the vector is odd under $\mathcal{P}$, the plane of rotation will have both axes flipped, which once again may be understood as a rotation through $\pi$. If the vector is even under either reflection, the commutation is trivial. Written in the same form as the above, we have:
\begin{equation}
\mathcal{P} R(\theta) = R(\theta)\mathcal{P}, \quad \mathcal{T} R(\theta) = R(\theta)\mathcal{T}.
\end{equation}
We say the rotation parameter $\theta$ is uncharged under the discrete reflection group: $K[\theta] = 1$. It is well known that any proper orthochronus Lorentz transform may be decomposed as a pure boost, followed by a rotation:
\begin{equation}
\Lambda(\beta, \theta) = R(\theta) B(\beta), \quad \forall \Lambda \in \operatorname{SO}^+(1,3),
\end{equation}
so we may condense the commutation relations:\footnote{Though not needed on a first read, appendix.~(\ref{sec:SemiLorentz}) continues directly from the current subsection.}
\begin{equation}
\label{secretsemi}
\mathcal{P} \Lambda(\theta, \beta) = \Lambda(\theta, -\beta)  \mathcal{P}, \quad \mathcal{T} \Lambda(\theta, \beta)  = \Lambda(\theta, -\beta) \mathcal{T}.
\end{equation}

%Outermorphism section to appendix
\vspace{-1.2em}
\subsection{\label{subsec:LTandPT} Constraining Reflection Charges with Lorentz Transforms}
\vspace{-0.4em}
The demand that vectors (and tensorial objects generally) represent covariant physical quantities is tantamount to the declaration that all inertial observers agree about the representations objects take under $\operatorname{O}(1,3)$, including reflections. Without this demand, naively each component of a vector could be assigned a distinct reflection charge leading to $4^4$ possible vector representations, most of which would be constructions which have preferred frames. As such the demand of covariance provides a much needed restriction on the kinds of objects which can enter our theories. Take a four vector $V^\mu$ in some inertial frame, with each Cartesian component having arbitrary charges $A,B,C,D$ under the discrete reflection group
\begin{equation}
K[V^\mu] = \begin{pmatrix} A\\
B\\
C\\
D\end{pmatrix}.
\end{equation}
Another frame rotated from the first by an angle $\theta$ in the $xy$ plane will see the vector with charges
\begin{equation}
K[(V^\mu)^\prime] = \begin{pmatrix} A\\
B\cdot K[\cos\theta] - C \cdot K[\sin\theta]\\
C \cdot K[\cos\theta] + B \cdot K[\sin\theta]\\
D\end{pmatrix}.
\end{equation}
As rotations themselves are uncharged under the reflection group, demanding that all observers related by a rotation agree about the image of the vector under any spacetime reflections, forces us to conclude that $K[B] = K[C] = K[D]$. All `spacial' components of a four vector will must have the same charge under the discrete reflection group. Now let us consider boosts acting upon vectors of the type
\begin{equation}
K[V^\mu] = \begin{pmatrix} A\\
\vec{B}\end{pmatrix},
\end{equation}
An observer boosted from the first frame will observe the vector as charged via
\begin{equation}
K[(V^\mu)^{\prime\prime}] = \begin{pmatrix} A \cdot K[\cosh\omega]  + B \cdot K[\sinh\omega]\\
B \cdot K[\cosh\omega] + A \cdot K[\sinh \omega] \end{pmatrix}.
\end{equation}
As previously discussed, the rapidity $\omega \propto v$ has reflection charge $PT$. As such, $K[\cosh\omega] = 1$ and $K[\sinh \omega] = PT$
which tells us the charges in the new frame are
\begin{equation}
K[(V^\mu)^{\prime\prime}] = \begin{pmatrix} A  + B \cdot PT\\
B + A \cdot PT \end{pmatrix}.
\end{equation}
For this transform to be covariant under the reflections, we must have that $B \cdot PT = A$. This give us only four kinds of four-vector which have well defined physical transformation properties for all observers related by $\operatorname{O}(1,3)$. These are:
\begin{equation}
\begin{pmatrix} T\\
\vec{P}\end{pmatrix}, \quad \begin{pmatrix} 1\\
\vec{PT}\end{pmatrix}, \quad \begin{pmatrix} P\\
\vec{T}\end{pmatrix}, \quad \begin{pmatrix} PT\\
\vec{1}\end{pmatrix}.
\end{equation}
We will name these respectively `coordinate type' (c), `momentum type' (m), `axial type' (a), and `polarization type' (p) four-vectors. When acting on these particular vector representations we will denote the operators with the shorthand $\mathcal{P}_{(c)}$, $\mathcal{T}_{(m)}$, etc, and utilize the notation $K[x^\mu] = (c)$, $K[p^\mu] = (m)$, etc. It is convenient to note that the basic operators here are all related up to a sign:
\begin{equation}
\begin{split}
&\mathcal{P}_{(c)} = \mathcal{P}_{(m)} = -\mathcal{P}_{(a)} = -\mathcal{P}_{(p)} \\
= -&\mathcal{T}_{(c)} \> = \mathcal{T}_{(m)} =\>\> \>\mathcal{T}_{(a)} \>= -\mathcal{T}_{(p)}.
\end{split}
\end{equation}
With vectors classified, we can begin to ask about the parity and time reversal of tensors generally.
\vspace{-0.6em}
\subsection{\label{subsec:2TensorReps}Simple 2-Tensor Representations of \texorpdfstring{$\operatorname{O}(1,3)$}{O(1,3)}}
\vspace{-0.6em}
%The set of reasonable assignments of reflection charges for four-vector components led us to only four possibilities, out of a naive count of sixteen. One is left to wonder if this pattern will continue, or if the proliferation of number of components will lead to greater numbers of allowed combinations of reflection charges. In Sec.~(\ref{FinalRepThry}) it is concluded there are always only four possible arrangements of reflection charges for tensors of any size.
\subsubsection{\label{subsubsec:LorGenRep}Generators of Lorentz Transforms}
%\vspace{-0.8em}
Consider the tensor product of two coordinate type vectors:
\begin{equation}
\label{eq:coordtens}
\begin{pmatrix}
T \\
\vec{P}
\end{pmatrix} \otimes \begin{pmatrix}
T \\
\vec{P}
\end{pmatrix} = \begin{pmatrix}
1 & \vec{PT} \\
\vec{PT} & 1 \mathbb{I}_3
\end{pmatrix}.
\end{equation}
This is the representation which the metric tensor lives in, since the metric components are defined as the inner product of coordinate basis vectors. From the information given in Sec.~(\ref{subsec:Boosts}) and standard representation theory, it can also be derived that the generators of Lorentz transforms upon vectors, live in this representation. The generators of boosts live in the $0i$ components of this kind of tensor, and the generators of rotations live in the $ij$ components. This agrees with our understanding that the generators of boosts and rotations are charged under discrete reflection group as $PT$ and $1$, respectively.

Define shorthand for a generic generator with boost parameters $\mathbf{K}$ and rotation parameters $\mathbf{L}$
\begin{equation}
g(\mathbf{K}, \mathbf{L}) = \begin{pmatrix}
0 & K_x & K_y & K_z\\
K_x & 0 & L_z & -L_y\\
K_y & -L_z & 0 & L_x\\
K_z & L_y & -L_x & 0
\end{pmatrix}.
\end{equation}
\noindent This tensor was constructed out of coordinate type vectors, so let us confirm that the operations of $\mathcal{P}$ and $\mathcal{T}$ acting in the defining representation behave as we anticipate. For time reversal we have
\begin{equation*}
\begin{split}
&\mathcal{T}\left[g(\mathbf{K}, \mathbf{L})\right] = \vspace{1ex} \mathcal{T}_{(c)} \>g(\mathbf{K}, \mathbf{L})\mathcal{T}_{(c)} \\
=&\vspace{1ex}\begin{pmatrix}
-1 & 0 & 0 & 0 \\
0 & 1 & 0 & 0 \\
0 & 0 & 1 & 0 \\
0 & 0 & 0 & 1 \\
\end{pmatrix}
\begin{pmatrix}
0 & K_x & K_y & K_z\\
K_x & 0 & L_z & -L_y\\
K_y & -L_z & 0 & L_x\\
K_z & L_y & -L_x & 0
\end{pmatrix}%\\
%&\qquad\qquad\qquad\qquad\qquad\qquad \times 
\begin{pmatrix}
-1 & 0 & 0 & 0 \\
0 & 1 & 0 & 0 \\
0 & 0 & 1 & 0 \\
0 & 0 & 0 & 1 \\
\end{pmatrix}
\end{split}
\end{equation*}
\vspace{-1.6em}
\begin{equation}
\hspace{-6em}\>= \begin{pmatrix}
0 & -K_x & -K_y & -K_z\\
-K_x & 0 & L_z & -L_y\\
-K_y & -L_z & 0 & L_x\\
-K_z & L_y & -L_x & 0
\end{pmatrix} = g(-\mathbf{K}, \mathbf{L}).
\end{equation} 
As for parity we have that $\mathcal{P}_{(c)} = -\mathcal{T}_{(c)}$, and so the signs can be made to cancel and the calculation gives the same: $\mathcal{P}\left[g(\mathbf{K}, \mathbf{L})\right] = g(-\mathbf{K}, \mathbf{L})$. We see both reflections are in agreement with the charges assigned in Eq.~($\ref{eq:coordtens}$). The cautious among us may be concerned that the generators of angular momentum have the same transformation properties under time reversal and parity, and that the transformation is trivial. Of course this is not a problem so long as the angular momentum itself has the correct transformation properties.
\vspace{-1em}
\subsubsection{\label{subsubsec:RelAngMom}Relativistic Angular Momentum}
%\vspace{-0.6em}
\noindent
Consider a relativistic angular momentum tensor $M$: 
\begin{equation}
\begin{split}
M^{\alpha \beta} &= x^\alpha p^\beta - x^\beta p^\alpha \\
\vspace{0.5ex}&=\begin{pmatrix}
0 & N_x & N_y & N_z\\
-N_x & 0 & J_z & -J_y\\
-N_y & -J_z & 0 & J_x\\
-N_z & J_y & -J_x & 0
\end{pmatrix} := M(\mathbf{N}, \mathbf{J}).
\end{split}
\end{equation}
Here $\mathbf{N}$ is the dynamic mass moment vector and $\mathbf{J}$ is the angular momentum vector. This tensor looks superficially identical to the previous example. However this tensor is built from vectors of different representations under the discrete reflection group, one coordinate type and one momentum type:
\begin{equation}
\begin{pmatrix}
T \\
\vec{P}
\end{pmatrix} \otimes \begin{pmatrix}
1 \\
\vec{PT}
\end{pmatrix} = \begin{pmatrix}
T & \vec{P} \\
\vec{P} & T \mathbb{I}_3
\end{pmatrix}.
\end{equation}
From the assignments of charges to the components we can see this is precisely what is expected: the angular momentum three vector will be odd under time reversal and even under parity. This case is the most interesting thus far, since we are combining vectors under different representations of $\operatorname{O}(1,3)$. Note the representation of $\mathcal{P}$ on $p^\mu$ and on $x^\mu$ is the same: $\mathcal{P}_{(c)} = \mathcal{P}_{(m)}$, and so the parity operator will act just as it did with the boost and rotation generators:
\begin{equation}
\mathcal{P}\left[M(\mathbf{N}, \mathbf{J})\right] = M(-\mathbf{N}, \mathbf{J}).
\end{equation}
On the other hand, the time reversal operator is represented differently on $p^\mu$ than it is on $x^\mu$, in particular $\mathcal{T}_{(m)} = -\mathcal{T}_{(c)}$. It turns out this does not introduce any headaches to the application of the operators. Firstly note it does not matter which side (more generally which index) the different representations of the time reversal operator act upon:
\begin{equation}
\mathcal{T}_{(c)} M \mathcal{T}_{(m)} = \left(-\mathcal{T}_{(m)}\right)  M \left(-\mathcal{T}_{(c)}\right) = \mathcal{T}_{(m)} M \mathcal{T}_{(c)}.
\end{equation}
Or much more explicitly treating the indices:

\vspace{-1em}
\begin{equation}
\begin{split}
\hspace{0.1em}&\mathcal{T}\left[M^{\alpha \beta}(\mathbf{N}, \mathbf{J})\right] \\
\hspace{-1em}=& (\mathcal{T}_{(c)} x)^\alpha (\mathcal{T}_{(m)} p)^\beta - (\mathcal{T}_{(c)} x)^\beta (\mathcal{T}_{(m)} p)^\alpha\\
\hspace{-1em}=& \left(\mathcal{T}_{(c)}\right)^{\alpha}_{\>\mu} x^\mu \left(\mathcal{T}_{(m)}\right)^{\beta}_{\>\nu} p^\nu- \left(\mathcal{T}_{(c)}\right)^{\beta}_{\>\nu} x^\nu \left(\mathcal{T}_{(m)}\right)^{\alpha}_{\>\mu} p^\mu\\
\hspace{-1em}=& \left(\mathcal{T}_{(c)}\right)^{\alpha}_{\>\mu} x^\mu \left(\mathcal{T}_{(m)}\right)^{\beta}_{\>\nu} p^\nu- \left(-\mathcal{T}_{(m)}\right)^{\beta}_{\>\nu} x^\nu \left(-\mathcal{T}_{(c)}\right)^{\alpha}_{\>\mu} p^\mu\\
\hspace{-1em}=& \left(\mathcal{T}_{(c)}\right)^{\alpha}_{\>\mu} x^\mu \left(\mathcal{T}_{(m)}\right)^{\beta}_{\>\nu} p^\nu- \left(\mathcal{T}_{(m)}\right)^{\beta}_{\>\nu} x^\nu \left(\mathcal{T}_{(c)}\right)^{\alpha}_{\>\mu} p^\mu\\
\hspace{-1em}=& \left(\mathcal{T}_{(c)}\right)^{\alpha}_{\>\mu} \left(\mathcal{T}_{(m)}\right)^{\beta}_{\>\nu} \left(x^\mu p^\nu- x^\nu p^\mu\right)\\
\hspace{-1em}=&\left(\mathcal{T}_{(c)}\right)^{\alpha}_{\>\mu} \left(\mathcal{T}_{(m)}\right)^{\beta}_{\>\nu} M^{\mu\nu} = \left(\mathcal{T}_{(m)}\right)^{\alpha}_{\>\mu} \left(\mathcal{T}_{(c)}\right)^{\beta}_{\>\nu} M^{\mu\nu}.
\end{split}
\end{equation}

This result shows that we may think of the tensor as having two indices which transform differently under the discrete reflection group, analogous to how co- and contravariant indices of the same tensor transform differently. However it is clear in this case the transform is ambivalent about `which' index transforms which way: we get the same result of the transform regardless. As such we may think of there being unambiguous tensor representations of the discrete reflection group, and we might say $K[M^{\alpha \beta}] = (cm)$, or (recall $g$ from above) $K[g] = (cc)$.\footnote{These `vector rep' indicators of tensor product reps are in fact not unique, since for example $(cc)$ and $(mm)$ are the same representation (as well as $(ca)$ and $(mp)$ etc.) A more satisfying representation theory of generic Lorentz tensors under $\operatorname{K}_4$ is presented in the Sec.~(\ref{FinalRepThry}). Regardless of the redundancy in this section, the important point is that these make the transformation properties of the tensor unambiguous.} From this, we know precisely which representation of the parity and time reversal operators to act upon the tensor with, and we know in fact the order in which they act does not matter. Now we may correctly and unambiguously perform the time reversal of an angular momentum tensor.
\begin{equation}
\begin{split}
    &\mathcal{T}\left[M(\mathbf{N}, \mathbf{J})\right] = \mathcal{T}_{c}\>M(\mathbf{N}, \mathbf{J})\mathcal{T}_{m}\\
\vspace{0.8em}=&\begin{pmatrix}
-1 & 0 & 0 & 0 \\
0 & 1 & 0 & 0 \\
0 & 0 & 1 & 0 \\
0 & 0 & 0 & 1 \\
\end{pmatrix}
\begin{pmatrix}
0 & N_x & N_y & N_z\\
-N_x & 0 & J_z & -J_y\\
-N_y & -J_z & 0 & J_x\\
-N_z & J_y & -J_x & 0
\end{pmatrix}\\
&\qquad\qquad\qquad\qquad\qquad \times 
\begin{pmatrix}
1 & 0 & 0 & 0 \\
0 & -1 & 0 & 0 \\
0 & 0 & -1 & 0 \\
0 & 0 & 0 & -1 \\
\end{pmatrix}\\
%\end{split}
%\end{equation}
%\begin{equation}
=&\begin{pmatrix}
0 & N_x & N_y & N_z\\
-N_x & 0 & -J_z & J_y\\
-N_y & J_z & 0 & -J_x\\
-N_z & -J_y & J_x & 0
\end{pmatrix} = M(\mathbf{N}, -\mathbf{J}).
\end{split}
\end{equation}
We can see application of the correct operators leads us to the standard and agreed upon time reversal transformation of angular momentum. 

\vspace{-1em}
\subsection{\label{FinalRepThry}Generic Tensor Representations of \texorpdfstring{$\operatorname{O}(1,3)$}{O(1,3)}}
\vspace{-0.2em}
We are now in a position to build up to a rather satisfying conclusion regarding generic tensor representations of $\operatorname{O}(1,3)$. We will see for every tensor power of the fundamental vector represenation of $\operatorname{SO}^+(1,3)$, the pattern of there being precisely four representations under $\operatorname{K}_4$ will persist: these can then be uniquely and consistently named so that the tensor product of generic representations simultaneously follows the representation theory of each $\operatorname{SO}^+(1,3)$ and $\operatorname{K}_4$.
\vspace{-1em}
\subsubsection{\label{subsec:scalars}Scalars and Vectors}
\label{subsec:Vectors}
The case of scalars under $\operatorname{O}(1,3)$ is already accounted for by the one dimensional irreducible representations of $\operatorname{K}_4$. They are what we have called $1, P, T, PT$. We have also seen there are precisely four well defined vector representations of $\operatorname{O}(1,3)$. Previously these were named $(c)$, $(m)$, $(a)$, and $(p)$. However one could imagine a more direct naming scheme: let us instead label our vector representations by the sign difference $\mathcal{P}$ and $\mathcal{T}$ take from the defining (coordinate) representation, given in Tab.~(\ref{tab:vecasK}). 

\begin{table}[htb]
\centering
\begin{tabular}{ c|cc } 
 & $\mathcal{P}$ & $\mathcal{T}$\\ 
 \hline
 $(c)$ & + & + \\ 
  $(m)$ & + & -- \\ 
  $(a)$ & -- & -- \\ 
  $(p)$ & -- & + \\ 
 \hline
\end{tabular}
\caption{A possible naming convention of the vector representations, indicating how their $\mathcal{P}$ and $\mathcal{T}$ operators differ from the familiar defining representation.}
\label{tab:vecasK}
\end{table} 

We may for example, denote the momentum representation as $\rho_{+-}$, and this would mean the parity operator acts simply as the defining representations parity operator, while the time reversal operator acts as the defining time reversal operator, with an additional sign flip, the intention being to forget about all the various incarnations of the operators, and only have to remember the defining ones, and then keep track of some signs. 

Curiously Tab.~(\ref{tab:vecasK}) resembles the $\mathcal{P}$ and $\mathcal{T}$ columns of Tab.~(\ref{K4reps}), the list of one dimensional representations of $\operatorname{K}_4$. In fact we could just as well label our scalar representations by their $\mathcal{P}$ and $\mathcal{T}$ parities just as well, identical to this renaming of the vector case. This seems to indicate we might be able to refer to momentum as being in something like the vector version of the $T$ representation, and this would tell us it transforms as in the defining representation, except that operators involving time reversal acquire an additional sign. We will confirm this below, that there are four representations of $\operatorname{O}(1,3)$ for every representation of $\operatorname{SO}^+(1,3)$, and these representations may be labeled by elements of the Klein group which in turn indicate additional sign changes garnered under the basic spacetime reflections.
\vspace{-1.5em}
\subsubsection{\label{subsec:Tensors}Tensors}
\vspace{-0.8em}
Employing the same logic as before, we may consider 2-tensor representations, and ask how these representation's parity and time reversal operations differ from application of those in the defining rep. By explicit computation, one can build Tab.~(\ref{2tensorreps}) of tensor products of vector representations. 
\begin{table}[h]
\centering
\begin{tabular}{ c|c|c|c|c } 
 $\otimes$ & $(c)$ & $(m)$ & $(a)$ & $(p)$\\ 
 \hline
 $(c)$ & $cc$ & $cm$ & $ca$ & $cp$\\ 
 \hline
 $(m)$ & $cm$ & $mm$ & $ma$ & $mp$ \\ 
 \hline
 $(a)$ & $ca$ & $ma$ & $aa$ & $ap$ \\ 
 \hline
 $(p)$ & $cp$ & $mp$ & $ap$ & $pp$ \\ 
 \hline
\end{tabular}
\quad = \quad
\begin{tabular}{ c|c|c|c|c } 
 $\otimes$ & $++$ & $+-$ & $--$ & $-+$\\ 
 \hline
 $++$ & $++$ & $+ -$ & $--$ & $-+$\\ 
 \hline
 $+-$ & $+ -$ & $++$ & $- +$ & $--$ \\ 
 \hline
 $--$ & $--$ & $- +$ & $++$ & $+ -$ \\ 
 \hline
 $-+$ & $- +$ & $--$ & $+-$ & $++$ \\ 
 \hline
\end{tabular}
\caption{The second tensor power of vector representations of $\operatorname{O}(1,3)$, uniquely identified by how their time and space inversion operations differ from tensors of the defining representation. }\vspace{-1em}
\label{2tensorreps}
\end{table}

One may have anticipated that higher tensor powers, by their having a greater number of components, would garner us a greater number of possible representations under reflections; this is false. We will always find exactly four representations of an $n^{\mathrm{th}}$ tensor power of $\operatorname{O}(1,3)$ reps, distinct in their spacetime reflection properties. The application of parity inversion on an $n^{\textrm{th}}$ order tensor will apply $n$ parity transformation matrices (of unspecified representation) contracting on the tensors $n$ indices. But these $n$ parity matrices can always be written as $n$ \textit{defining} parity transformation matrices, and a possibly left over sign. Turning this around we may say parity inversion of any $n^{\textrm{th}}$ order tensor \textit{is} the application of $n$ defining transforms, and the possible application of a sign. An identical argument holds for $\mathcal{T}$. 

Taking a look at Tab.~(\ref{2tensorreps}), hidden in plain sight is the multiplication table of $\operatorname{K}_4$, our discrete reflection group! It becomes explicit if we perform the identification of names of representations given in Tab.~(\ref{tab:signAsK4}).

\begin{table}[!t]
\centering
\begin{tabular}{ c|c} 
 Sign change of $\mathcal{P}$ and $\mathcal{T}$ from defining & Analogous Rep of $\operatorname{K}_4$ \\ 
 \hline
 ++ (Coordinate Type)& $1$ \\ 
  +-- (Momentum Type)& $T$ \\ 
  \hspace{-1.5em}-- -- \hspace{1em}(Axial Type)& $PT$\\ 
  \hspace{0.7em}-- + (Polarization Type) & $P$ \\ 
 \hline
\end{tabular}
\caption{Generic $\operatorname{O}(1,3)$ tensors show up in `reflection quartets' which can be labeled just as the one dimensional representations. The details of the reflections are still more complicated than the one dimensional case, but the tensor products will follow the same pattern of $\operatorname{K}_4$.}\vspace{-1.5em}
\label{tab:signAsK4}
\end{table}

It is easy to see the right side version of Tab.~(\ref{2tensorreps}) remains true no matter the underlying representation of $\operatorname{SO}^+(1,3)$, and so every representation of $\operatorname{SO}^+(1,3)$ will have four corresponding representations in $\operatorname{O}(1,3)$, one of each type $1, P, T, PT$. We see from this, for example, that a tensor product of a coordinate type vector (`$1$ type vector') with a momentum type vector (`$T$ type vector'), will give a `$T$ type second order tensor', i.e. this type of tensor transforms as the defining rep, but under any time reversal the components acquire an additional overall sign. This enormously simplifies keeping track of the transformation properties of tensors components: it is in principle no more complicated than the $1$ dimensional representations. This assignment of names agrees with the scalar case: $x^\mu p_\mu$ is a scalar which is odd under $\mathcal{T}$, and it takes the scalar representation $1 \cdot T = T$. 

It bears repeating that when saying a tensor or vector takes a representation `$T$' or `$PT$' etc, we are not merely saying that the whole tensor picks up a sign under those transforms. We are saying that \textbf{when `naively' applying the \textit{defining} parity or time reversal} matrices to the object, \textbf{those particular transforms must pick up an additional sign in order actually be the correct transform} of that object. Mathematically of course this is the same as treating the representation theory as if there are only $\mathcal{P}_\mathrm{def}$ and $\mathcal{T}_\mathrm{def}$, and as if the (for e.g.) $T$ charge simply acts as in the one dimensional case picking up a sign under time reversals. The appeal of this mode of thinking is that we may largely forget about all the ways $\mathcal{P}$ and $\mathcal{T}$ can act on the representations --- we need only remember the defining reps, and the $1$-dimensional representations of the Klein group. However it is important to keep in mind that the seemingly `extra' signs are absolutely necessary to correct for using the operators from the wrong representation; the apparently `extra' signs do not enter via an ad hoc extension. 
\vspace{-2.5em}
\subsection{\label{subsec:Examples}Important Examples}
\vspace{-1em}
\subsubsection{\label{subsubsec:RaisingLoweringIndices}The Metric: Raising and Lowering Indices}
\vspace{-0.6em}
It is worth seeing why $1$-type tensors deserve the name. The metric tensor is a $1$-type tensor: this is reassuring since we do not anticipate that the parity and time reversal properties should be affected by the raising and lowering of indices. We can see this for example with a $T$ type tensor, such as the Faraday or angular momentum tensors. One can readily verify, in terms of the one dimensional representations of the components:
\begin{equation}
\begin{split}
&\begin{pmatrix}
1 & \vec{PT} \\
\vec{PT} & 1 \mathbb{I}_3
\end{pmatrix} \begin{pmatrix}
T & \vec{P} \\
\vec{P} & T \mathbb{I}_3
\end{pmatrix} \begin{pmatrix}
1 & \vec{PT} \\
\vec{PT} & 1 \mathbb{I}_3
\end{pmatrix} \\
=&\begin{pmatrix}
1 & \vec{PT} \\
\vec{PT} & 1 \mathbb{I}_3
\end{pmatrix} \begin{pmatrix}
T & \vec{P} \\
\vec{P} & T \mathbb{I}_3
\end{pmatrix} = \begin{pmatrix}
T & \vec{P} \\
\vec{P} & T \mathbb{I}_3
\end{pmatrix}.
\end{split}
\end{equation}
This is an example of how the multiplication of the Klein group representations (here $1 \cdot T \cdot 1 = T$) may be carried out no matter the underlying $\operatorname{SO}^+(1,3)$ structure of the tensors involved. Interaction with $1$-type tensors in general will leave us in the same type of reflection representation.

\vspace{-1em}
\subsubsection{\label{subsubsec:LeviCivita}The Levi-Civita Symbol \texorpdfstring{\&}{&} Pauli–Lubanski Pseudo-Vector }
%\vspace{-0.2em}
The Pauli-Lubanski pseudo-vector is defined as follows, given an angular momentum tensor $M$ and a momentum vector $p$:
\begin{equation}
W_{\mu} = \frac{1}{2} \varepsilon_{\mu\nu\sigma \rho} M^{\nu \sigma} p^\rho = 
\begin{pmatrix}
\mathbf{J}\cdot \mathbf{p} \\
E \mathbf{J} - \mathbf{p} \times \mathbf{N}
\end{pmatrix}.
\end{equation}
Typically this object is most useful when $M$ and $p$ are taken to be operators, and values of the scalar $W^\mu W_\mu$ are used to classify different representations of the Poincare group \cite{Weinberg_1995}. Here we have two objects for which we do not know their transformation properties under the reflection group, $W$ and $\varepsilon$. However, we understand the transformation properties of $M$ and $p$ and so after computation can figure out by inspection the transformation properties of the components of $W$, which can then be used to tell us about the Levi-Civita symbol. The first component, $W_0$, is odd under parity inversion, and overall even under time reversal, as both $\mathbf{J}$ and $\mathbf{p}$ are $\mathcal{T}$-odd. This means $W_0$ is charged under the reflection group as $P$ --- by our classification of vectors, we expect the spacial part of $W$ to be charged as $T$. Inspection shows that is true. So we have confirmed that $W^\mu$ is an axial type vector, i.e a $PT$ type vector, and will `pick up a sign' under each of time and space reflections. 

From this, we can now deduce the charge of $\varepsilon_{\mu\nu\sigma \rho}$ under the reflection group. With axial vectors as `PT-type' vectors, we can deduce
\begin{equation}
\begin{split}
K[W_{\mu}] &= K[\varepsilon_{\mu\nu\sigma \rho}] K[M^{\nu \sigma}] K[ p^\rho ]\\
PT &= K[\varepsilon_{\mu\nu\sigma \rho}] \cdot T \cdot T\\
PT &= K[\varepsilon_{\mu\nu\sigma \rho}].
\end{split}
\end{equation}
The Levi-Civita symbol transforms just as any four index tensor, however it picks up signs under space or time reflections. This may not be the fact it appears to you as on the surface: as $\varepsilon^{\mu\nu\sigma \rho}$ is the invariant tensor which delivers us determinants, for which the reflections $\mathcal{P}_\mathrm{def}$ and $\mathcal{T}_\mathrm{def}$ have $-1$, the charge under the reflection group will cancel these signs out. Thus it is the case that in fact $\varepsilon$ is exactly invariant under the entire Lorentz group.\footnote{It is worth mentioning this fact about $\varepsilon$ is typically understood via the concept of tensor densities (see \cite{Weinberg_1972}). I believe this purely representation theoretic perspective gives a much more satisfying understanding of the Levi-Civita symbol's properties. }
\vspace{-1em}
\section*{\label{sec:Summary}Summary and Conclusions}
\vspace{-0.5em}
A convenient representation theory of the entire Lorentz group $\operatorname{O}(1,3)$ has been constructed. In particular the decomposition $\operatorname{O}(1,3) = \operatorname{SO}^+(1,3) \rtimes \operatorname{K}_4$, understood as the fact that boost parameters are charged under parity and time reversal, constrains the representations of the discrete reflection group $\operatorname{K}_4$ upon four-vectors to be one of four combinations, and subsequently it was shown there are precisely four distinct representations of $\operatorname{O}(1,3)$ for each representation of $\operatorname{SO}^+(1,3)$. These four distinct representations can be labeled $1, P, T,$ and $PT$; and the label of the representation tells us precisely the tensor's behavior under spacetime reflections $\mathcal{P}$ and $\mathcal{T}$. Specifically, all tensors may be transformed by the defining (coordinate) representation of the parity and time reversal operators, and those in the $P,T$, or $PT$ representations gain an additional sign under the transforms in their names. 

The representation of a product of any two tensors, whether through an inner product, tensor product, contraction etc, will be labeled by the multiplication of the $\operatorname{K}_4$ labels of each factor, completely agnostic to the particular $\operatorname{SO}^+(1,3)$ structure of the tensors at play. So for example the contraction of a $PT$ type 2-tensor with a $T$ type vector results in a $P$ type vector. In toto, in this formalism the representation of say, the Faraday tensor $F^{\mu\nu} = x^\mu p^\nu - x^\nu p^{\mu}$ in the full group $\operatorname{O}(1,3)$ would be given as
\begin{equation}
\left[\left(\frac{1}{2},\frac{1}{2}\right)\wedge \left(\frac{1}{2},\frac{1}{2}\right), 1\cdot T\right] = \left[(1,0)\oplus(0,1), T\right].
\end{equation}
Since representations of $\operatorname{SO}^+(1,3)$ can be labeled by pairs of half integers, representations of $\operatorname{O}(1,3)$ can be labeled by pairs of half integers, alongside a charge under the discrete reflection group $\operatorname{K}_4$, interpreted as above. Several further examples are given Appendix.~(\ref{sec:examples}) to show off the utility of the formalism. By keeping track of a simple label the behavior of any tensorial object under arbitrary reflections is made transparent to us. 
\vspace{-1em}
\subsection*{Future Work}
\vspace{-1em}
The author intends to pursue a similar project for the spinoral double coverings of $\operatorname{O}(1,3)$, the $\operatorname{Pin}$ groups, in order to further study, classify, and clarify the relationship between Lorentz invariance, parity violation, and the $\mathcal{CPT}$ theorem.
\vspace{-1em}
\section*{Author Declarations}
\vspace{-1em}
The author has no conflicts to disclose. Data sharing is not applicable to this article as no new data were created or analyzed in this study.

\newpage
\appendix
\section{Further Examples}
\label{sec:examples}
\subsection{\label{subsubsec:Amu}The Faraday Tensor and the Potential \texorpdfstring{$A^\mu$}{A}}
\vspace{-0.6em}
While the four potential $A^\mu$ is not a 2-tensor, a discussion of the Faraday tensor is incomplete without first addressing it. The parity inversion properties of $A^\mu$ are rarely up for debate. It is the time reversal of $A^\mu$ which is the cause of contention in the literature. In particular, it is the question of whether a time reversal should `include' a charge conjugation. This debate is due in part to the Feynman–Stückelberg interpretation of anti-particle solutions, and in part due to the anti-unitarity of the time reversal operator. However, so long as we are considering the case of classical fields, we can cleanly separate time reversal from charge conjugation without debate. In what follows, the time reversal of a stationary charge $q$ will be understood to produce a description of an identical system, and \textit{not} a description of a system of a stationary charge of $-q$. This can be stated as the postulate that classical electrical charges live in the trivial representation of the discrete reflection group. 

It is not clear \textit{a priori} which vector rep of $\operatorname{O}(1,3)$ the potential $A^\mu$ takes. One could argue from its equations of motion that it should have the same properties as a four-current; but this requires first assuming parity and time reversal invariance of those equations. The answer can be gleaned from the much less assumptive but important expectation that the canonical momentum of the electromagnetic field has well defined transformation properties under the discrete reflection group for all observers. The canonical momentum which enters the Hamiltonian of the electromagnetic field is given by \cite{JACKSON3RD}:
\begin{equation}
P^\mu = \left(p^\mu + e A^\mu\right),
\end{equation}
where $p^\mu$ is the mechanical momentum of the relevant particle with charge $e$. If observers related by (proper orthochronus) Lorentz transforms are to agree about the charges of $P^\mu$ under the discrete reflection group, it must be the case that $K[e A^\mu] =K[p^\mu] = (m)$. Committing to our assumption that $e$ is invariant under the reflection group, we must have $K[A^\mu] = K[p^\mu] = 1 \oplus \vec{PT} = (m)$, i.e. $A^\mu$ is a momentum type vector. 

We are now in a place to simply understand the time reversal of the Electromagnetic fields. The Faraday tensor may be defined as 
\begin{equation}
\begin{split}
F^{\mu\nu}(\mathbf{E}, \mathbf{B}) &= \partial^\mu A^\nu - \partial^\nu A^\mu \\
&= \begin{pmatrix}
0 & -E_x & -E_y & -E_z\\
E_x & 0 & -B_z & B_y\\
E_y & B_z & 0 & -B_x\\
E_z & -B_y & B_x & 0
\end{pmatrix}.
\end{split}
\end{equation}
This representation of $\operatorname{O}(1,3)$ is identical to that of the angular momentum tensor:
\begin{equation}
K[F^{\mu\nu}] = \begin{pmatrix}
T \\
\vec{P}
\end{pmatrix} \otimes \begin{pmatrix}
1 \\
\vec{PT}
\end{pmatrix} = \begin{pmatrix}
T & \vec{P} \\
\vec{P} & T \mathbb{I}_3
\end{pmatrix}.
\end{equation}
As such, parity and time reversal will act the same here as they did upon the angular momentum tensor:
\begin{equation}
\begin{split}
\mathcal{P}\left[F(\mathbf{E}, \mathbf{B})\right] = F(-\mathbf{E}, \mathbf{B}),\\
\mathcal{T}\left[F(\mathbf{E}, \mathbf{B})\right] = F(\mathbf{E}, -\mathbf{B}).
\end{split}
\end{equation}
This is what is taught in most physics textbooks, and finally we can see a consistent representation theoretic defense of it. The only room for disagreement is the debate of whether or not charges should take non-trivial representations of the discrete reflection group, which presently seems a discussion better suited for realm of quantum field theory. We may also see clearly why what Arntzenius \& Greaves \cite{GREAVESARNTZENIUS} dubbed `Malaments proposal' works:
\begin{equation}
\begin{split}
\mathcal{T}_{c}\>F(\mathbf{E}, \mathbf{B})\mathcal{T}_{m} &= \left(-\mathcal{P}_{(c)}\right)F(\mathbf{E}, \mathbf{B})\left(\mathcal{P}_{(m)}\right)\\
&= -F (-\mathbf{E}, \mathbf{B})\\
&= F (\mathbf{E}, -\mathbf{B}).
\end{split}
\end{equation}
It is clear why the time reversal of the Faraday tensor should look like the application of parity, followed by an overall additional minus sign on the whole tensor.
\vspace{-1em}
\subsection{\label{subsubsec:Efield}An Observers \texorpdfstring{$\mathbf{E}$}{Electric} Field}
\vspace{-0.6em}
Given the Faraday tensor $F^{\mu\nu}(\mathbf{E}, \mathbf{B})$ in one frame, an observer in another frame with relative four velocity $u^\alpha$ will observe the electric field to be:
\begin{equation}
E^\mu = F^{\mu\nu} u_\nu,
\end{equation}
and the combination of charges under the reflection group gives us:
\begin{equation}
K[E^\mu] = K[F] \cdot K[u] = T \cdot T = 1.
\end{equation}
Without digging into the components, we already know the resulting vector is a $1$-type vector, a.k.a. a coordinate type vector, and so its parity and time reversal properties are transparent to us without the need to pick apart its various pieces. For reference, in the rest frame this simply gives: $E^\mu = (0 \>\> \mathbf{E})^T$, which being a coordinate type vector solidifies our understanding that electric fields are odd under parity and even under time reversal.
\vspace{-1em}
\subsection{\label{subsubsec:Bfield}An Observers \texorpdfstring{$\mathbf{B}$}{Magnetic} Field}
\vspace{-0.6em}
An observer's $\mathbf{B}$ field is defined analogously to an observers $\mathbf{E}$ field, but with the (Hodge) dual of the Faraday tensor. For $F$ in our frame and an observer with relative four velocity $u^\nu$, we have that:
\begin{equation}
B^\mu = \widetilde{F}^{\mu\nu} u_{\nu} = \frac{1}{2} \varepsilon^{\mu\nu\gamma\beta}F_{\gamma\beta} u_{\nu}.
\end{equation}
The combination of charges under the reflection group gives us:
\begin{equation}
\begin{split}
K[B^\mu] &= K[\tilde{F}] \cdot K[u] \\
&= K[\varepsilon] \cdot K[F] \cdot K[u]\\
&= PT \cdot T\cdot T = PT,
\end{split}
\end{equation}
and so we find the observers $\mathbf{B}$ field is an axial type vector, which once again agrees with our prior understanding. Of course, in the rest frame we have $B^\mu = (0 \>\> \mathbf{B})^T$.
\vspace{-2.1em}
\subsection{\label{subsubsec:Maxwell}Maxwell's Equations}
\vspace{-0.6em}

Just as we can be sure of a tensor equation's covariance by knowing each term transforms covariantly (each term transforms as a tensor, and as the same kind of tensor),\footnote{It is important to note that in general this is not the only way for things to covariant. For example Christoffel symbols on their own on are not covariant objects, but when combined with a partial derivative, the combination is covariant.} we might ask of an analogous criteria for $\operatorname{O}(1,3)$. For familiarity and to be explicit, below are Maxwell's equations in three-vector notation:
\begin{equation}
\label{eq:maxwell}
\begin{split}
\vec{\nabla} \cdot \mathbf{E}(\mathbf{x}, t) &= \rho(\mathbf{x}, t), \\
\vec{\nabla} \cdot \mathbf{B}(\mathbf{x}, t) &= 0, \\
\vec{\nabla} \times \mathbf{E}(\mathbf{x}, t) &= -\frac{1}{c}\frac{\partial \mathbf{B}(\mathbf{x}, t)}{\partial t}, \\
\vec{\nabla} \times \mathbf{B}(\mathbf{x}, t) &= \frac{1}{c} \mathbf{J}(\mathbf{x}, t) +\frac{1}{c}\frac{\partial \mathbf{E}(\mathbf{x}, t)}{\partial t} .
\end{split}
\end{equation}
Noting the reflection charges of each object:
\begin{equation}
\begin{split}
K\left[\frac{\partial}{\partial t}\right] &= T, \quad K[\rho] = 1, \quad K[\mathbf{E}] = P,\\
K[\nabla] &= P, \quad K[\mathbf{J}] = PT, \quad K[\mathbf{B}] = T. \\
\end{split}
\end{equation}
We can see by inspection each individual Maxwell equation has only terms with a single well defined reflection charge. In order: $1, PT, 1, PT$. That an equation is constructed out of terms which all share the same charge under the reflection group will be known as being `Reflection Homogeneous'. It can be seen that reflection covariance of equations of motion implies that spacetime-reflected solutions are also solutions. Reflection covariance of a theory implies parity and time reversal invariance of a theory, it is the analogue of manifest covariance. Let us inspect the equations of motion for the parity and time reflected solutions. For both cases it is useful to note:
\begin{equation}
\frac{\partial}{\partial t} = - \frac{\partial}{\partial (-t)}, \quad \vec{\nabla}_{\mathbf{x}} = - \vec{\nabla}_{(-\mathbf{x})}.
\end{equation}
\vspace{-2em}
\subsubsection{\label{subsubsec:ParityEM}Parity Symmetry}
%\vspace{-0.2em}
Given fields $\mathbf{E}(\mathbf{x},t)$ and $\mathbf{B}(\mathbf{x},t)$ which are solutions to Maxwell's equations~(\ref{eq:maxwell}), we may inspect the same structure of the parity transform of the solutions:
\begin{equation}
\begin{split}
\mathbf{E}_\mathcal{P}(\mathbf{x}, t) &:= \mathcal{P}(\mathbf{E}(\mathbf{x},t)) = -\mathbf{E}(-\mathbf{x},t),\\
\mathbf{B}_\mathcal{P}(\mathbf{x}, t) &:= \mathcal{P}(\mathbf{B}(\mathbf{x},t)) = \mathbf{B}(-\mathbf{x},t).
\end{split}
\end{equation}
What we are inspecting is whether or not the parity inverted solution, understood as a possible configuration of fields in our world, is also a solution to Maxwell's equations. I.e. we should think of $\mathbf{E}_\mathcal{P}$ precisely as some new vector valued function of our coordinates $\mathbf{x}$, and ask if this new function is also a solution. Let us show that it is, equation by equation. 
\begin{equation}
\begin{split}
\vec{\nabla}_{\mathbf{x}} \cdot \mathbf{E}_\mathcal{P}(\mathbf{x}, t) &= \vec{\nabla}_{\mathbf{x}} \cdot \left(-\mathbf{E}(-\mathbf{x}, t)\right) \\
&= \vec{\nabla}_{(-\mathbf{x})} \cdot \left(\mathbf{E}(-\mathbf{x}, t)\right)\\
&= \rho(-\mathbf{x}, t) \\
&= \rho_{\mathcal{P}}(\mathbf{x}, t).
\end{split}
\end{equation}
Thus the parity inverted electric field satisfies Gausses law if the original field does. Next we have
\begin{equation}
\begin{split}
\vec{\nabla}_{\mathbf{x}} \cdot \mathbf{B}_\mathcal{P}(\mathbf{x}, t) &= -\vec{\nabla}_{(-\mathbf{x})} \cdot \left(\mathbf{B}(-\mathbf{x}, t)\right) = 0.
\end{split}
\end{equation}
Similarly if the original equation yields no magnetic charges, neither will the parity inverted solution. Inspecting the third of Maxwell's equations we have
\begin{equation}
\begin{split}
\vec{\nabla}_{\mathbf{x}} \times \mathbf{E}_\mathcal{P}(\mathbf{x}, t) &= \vec{\nabla}_{(-\mathbf{x})} \times \left(\mathbf{E}(-\mathbf{x}, t)\right) \\
&= -\frac{1}{c} \frac{\partial \mathbf{B}(-\mathbf{x}, t)}{\partial t} \\
&= -\frac{1}{c} \frac{\partial \mathbf{B}_\mathcal{P}(\mathbf{x}, t)}{\partial t}.
\end{split}
\end{equation}
Which shows the same result for Faraday's law. Finally Ampere's law
\begin{equation}
\begin{split}
\vec{\nabla}_{\mathbf{x}} \times \mathbf{B}_\mathcal{P}(\mathbf{x}, t) &= -\vec{\nabla}_{(-\mathbf{x})} \times \mathbf{B}(-\mathbf{x}, t) \\
&=- \frac{1}{c} \mathbf{j}(\mathbf{-x},t)-\frac{1}{c} \frac{\partial \mathbf{E}(-\mathbf{x}, t)}{\partial t} \\
&= \frac{1}{c} \mathbf{j}_\mathcal{P}(\mathbf{x},t) + \frac{1}{c} \frac{\partial \mathbf{E}_\mathcal{P}(\mathbf{x}, t)}{\partial t}.
\end{split}
\end{equation}
Thus the parity inversion of any solution will yield an additional solution. This is what is meant by declaring classical electromagnetism respects parity.
\vspace{-0.8em}
\subsubsection{\label{subsubsec:TimeEM}Time Reversal Symmetry}
%\vspace{-0.2em}
Again, given fields which are solutions to Maxwell's equations, recall the time reversed fields are given by
\begin{equation}
\begin{split}
\mathbf{E}_\mathcal{T}(\mathbf{x}, t) &:= \mathcal{T}(\mathbf{E}(\mathbf{x},t)) = \mathbf{E}(\mathbf{x},-t), \\
\mathbf{B}_\mathcal{T}(\mathbf{x}, t) &:= \mathcal{T}(\mathbf{B}(\mathbf{x},t)) = -\mathbf{B}(\mathbf{x},-t).
\end{split}
\end{equation}
In this case the coordinate dependence of $\nabla$ will be repressed with the understanding all spacial derivatives are with respect to $\mathbf{x}$. The first two equations are simple in this case, as they are constraints at all times $t$ and not dynamical
\begin{equation}
\begin{split}
\vec{\nabla} \cdot \mathbf{E}_\mathcal{T}(\mathbf{x}, t) &= \vec{\nabla} \cdot \mathbf{E}(\mathbf{x}, -t) = \rho(\mathbf{x}, -t) = \rho_{\mathcal{T}}(\mathbf{x}, t),\\
\vec{\nabla} \cdot \mathbf{B}_\mathcal{T}(\mathbf{x}, t) &= \vec{\nabla} \cdot \left(\mathbf{B}(\mathbf{x}, -t)\right) = 0.
\end{split}
\end{equation}
So if the initial fields satisfy Gauss' law, so too will the time reversed fields. As for Faraday's law we have
\begin{equation}
\begin{split}
-\frac{1}{c}\frac{\partial \mathbf{B}_\mathcal{T}(\mathbf{x},t)}{\partial t} &= -\frac{1}{c}\frac{\partial \mathbf{B}(\mathbf{x},-t)}{\partial (-t)} \\
&= \vec{\nabla} \times \mathbf{E}(\mathbf{x},-t)\\
&= \vec{\nabla} \times \mathbf{E}_\mathcal{T}(\mathbf{x},t)
\end{split}
\end{equation}
and lastly for Ampere's law
\begin{equation}
\begin{split}
\frac{1}{c}\frac{\partial \mathbf{E}_\mathcal{T}(\mathbf{x},t)}{\partial t} &= -\frac{1}{c}\frac{\partial \mathbf{E}(\mathbf{x},-t)}{\partial (-t)}\\
&= -\left(- \frac{1}{c} \mathbf{j}(\mathbf{x},-t) +\vec{\nabla} \times \mathbf{B}(\mathbf{x},-t) \right)\\
&= - \frac{1}{c} \mathbf{j}_\mathcal{T}(\mathbf{x},t) + \vec{\nabla} \times \mathbf{B}_\mathcal{T}(\mathbf{x},t).
\end{split}
\end{equation}
Therefore if we have solutions to Maxwell's equations, the time reversal of those solutions will also be solutions. 

\vspace{-2em}
\subsection{\label{subsubsec:Heat}The Heat Equation}
\vspace{-0.6em}
\label{heateq}
Although we found Maxwell's equations to be Parity and Time reversal invariant, it did not `have' to be the case. By this I mean we did not find the reflection charges by demanding this property, it came out as a result. A contrasting example is the heat equation, given by 
\begin{equation}
\nabla^2 u = \frac{\partial u}{\partial t},
\end{equation}
for $u(\mathbf{x},t)$ a scalar function. Let us demand the equation be reflection homogeneous and see what falls out:
\begin{equation}
\begin{split}
K[\nabla^2 u] &= K\left[\frac{\partial u}{\partial t}\right] \\
K[\nabla^2] K[u] &= K\left[\frac{\partial }{\partial t}\right] K[u]\\
1 \cdot K[u] &= T \cdot K[u].\\
\end{split}
\end{equation}
We immediately find a problem. There are no possible assignments of $K[u]$ which could make this equation true. While both sides will necessarily have the same properties under parity inversion, the two sides have clearly different time reversal properties, even \textit{if} we are free to choose $K[u]$. Thus for a solution $u$ to the heat equation, both $u_\mathcal{P}(\mathbf{x},t) = \pm u(-\mathbf{x},t)$ will also be solutions to the heat equation, whereas time reversed solutions $u_\mathcal{T}(\mathbf{x},t) = \pm u(\mathbf{x},-t)$ cannot be. So for any system obeying the heat equation, there is no physical `reversed' setup which will see the opposite behavior of the system, and as such the behavior of these systems must be irreversible.

\section{\texorpdfstring{$\operatorname{O}(1,3)$}{O(1,3)} as a Semi-direct Product Group}
\vspace{-0.6em}
\subsection{Semi-Direct Products}
\vspace{-0.6em}
\label{SemiDirect}
Given two groups $G$ and $H$, and the automorphism group of $G$: $\operatorname{Aut(G)}$,\footnote{The automorphism group of any group, is set of all bijective group homomorphisms: $G \rightarrow G$ \cite{AbsAlg}.} let us suppose there is a way to map elements of $H$ to elements of $\operatorname{Aut(G)}$, and call this map $\psi$. I.e. for every element $h \in H$, $\psi(h) := \psi_h$ is an automorphism of $G$. Given this map there is now a way for elements of the group $H$ to act on elements of the group $G$: 
\begin{equation}
h \cdot g = \psi_h(g).
\end{equation}
A semi-direct product of the two groups $G$ and $H$, necessitates a choice of $\psi$, and is denoted by $G \rtimes_\psi H$ (usually the dependence on $\psi$ is omitted and implicit if there is a natural action of $H$ on $G$). Once $\psi$ is understood, i.e. once there is a well defined action of $H$ upon $G$, then we may define a semi-direct product. 

Just as with the direct product, elements of a semi-direct product group are ordered pairs of elements of the underlying groups
\begin{equation}
(g,h) \in G \rtimes H.
\end{equation}
It is in the multiplication of group elements of a semi-direct product that $\psi$ enters. For two elements of a semi-direct product group, the group multiplication law is the following:
\begin{equation}
\begin{split}
(g_2, h_2) \cdot (g_1, h_1) &= (g_2 \cdot \left(h_2 \cdot g_1 \right), h_2 \cdot h_1 ) \\
&= (g_2 \cdot \psi_{h_2}(g_1), h_2 \cdot h_1 ) .
\end{split}
\end{equation}
This looks mostly like the law for a direct product of group actions, except that the group multiplication for the $G$ component has been `twisted' by $\psi$. This may be interpreted as a commutation relation: the effect of moving $h_2$ `past' $g_1$ results in $g_1$ being altered in the way defined by $\psi$. It was seen above that this behavior is precisely how the discrete reflection group $\operatorname{K}_4$ interacts with the proper orthochronus Lorentz group. 

\vspace{-1em}
\subsection{\label{sec:AutLor}Outer Automorphisms of the Lorentz Group}
\vspace{-0.6em}
All group automorphisms may be classified as being either inner or outer. The \textit{inner} automorphisms of the Lorentz group are well understood: these are given by the adjoint action of the Lorentz group on itself, i.e. conjugation of a (proper orthochronus) Lorentz transform by any other. These relate the Lorentz transforms of one observer, to the corresponding Lorentz transforms of another observer at the same point. These automorphisms are not particularly important for the discussion at hand, and it is the \textit{outer} automorphism group which will play the crucial role in our semi-direct product. Outer automorphisms are any automorphisms which are not inner.

The Lie algebra of the Lorentz group, $\mathfrak{spin}(1,3, \mathbb{R})$ is known to have one non-trivial (outer) automorphism. Another way to say this is that the algebra's outer automorphism group is $\mathbb{Z}_2$.\footnote{For those familiar with the study of semi-simple Lie groups generally, this is merely a statement about the Dynkin diagram of the Lorentz Lie algebra. The symmetry can be understood as `swapping' the nodes of the diagram $D_2 \cong A_1 \times A_1$.\begin{equation*} D_2 = \begin{array}{l}
\dynkin{A}{1}\\
\dynkin{A}{1}
\end{array}\end{equation*}
}
To see this explicitly, if we define the Lie algebra via the standard Cartesian basis of generators of rotations ($L_i$) and boosts ($K_i$):
\begin{equation}
\begin{split}
\left[L_i, L_j\right] &= \varepsilon_{ijk} L_k,\\
\left[L_i, K_j\right] &= \varepsilon_{ijk} K_k,\\ 
\left[K_i, K_j\right] &= -\varepsilon_{ijk} L_k,
\end{split}
\end{equation}
then the two fold symmetry which is our outer automorphism, takes the concrete form of a `reflection' of the algebra:
\begin{equation}
L_i \mapsto L_i, \quad K_i \mapsto -K_i.
\end{equation}
That is, if in this six dimensional algebra we reflect the three boost directions (i.e. redefine the boost generators to be their negations), the algebra remains unchanged. We can be sure this is not an inner automorphism, since the automorphism has a determinant of $-1$, which no Lorentz transform in the proper orthochronus subgroup has. It is well known that for any Lie algebra homomorphism there exists a corresponding Lie group homomorphism,\footnote{The derivative of the group homomorphism is the algebra homomorphism.} so we will refer to this outer automorphism, this `boost reflection', by the generic name $\mathcal{\hat{O}}$ regardless of whether we think of it as acting on the group or the algebra. Note with this we may write our outer automorphism group as $\{1, \mathcal{\hat{O}}\}$.\footnote{One interesting consequence of this symmetry is that representations of the Lorentz group generically should be expected to come in `pairs' related by this automorphism. This is why we find left and right handed spinors, self-dual and anti-self-dual field strength tensors, etc.}

\vspace{-1em}
\subsection{\label{sec:SemiLorentz}\texorpdfstring{Semi-Direct Product Structure of $\operatorname{O}(1,3)$}{Semi-Direct Product Structure of O(1,3)}}
\vspace{-0.6em}

On a generic Lorentz transform, $\mathcal{P}$ and $\mathcal{T}$ leave the rotation angles the same, but reverse the boost parameters. This action is equivalent to saying we have acted via an outer automorphism $\mathcal{\hat{O}}$ upon our Lie group:\footnote{See appendix~(\ref{sec:AutLor}) for review of the outer automorphism group of the proper orthochronus Lorentz group}
\begin{equation}
\Lambda(\theta, -\beta) = \mathcal{\hat{O}}(\Lambda(\theta, \beta)).
\end{equation}
With this in mind we can rewrite Eq.~($\ref{secretsemi}$) in a much more suggestive fashion:
\begin{equation}
\label{obvsemi}
\begin{split}
\mathcal{P} \Lambda(\theta, \beta) &= \mathcal{\hat{O}}\left(\Lambda(\theta, -\beta)\right)  \mathcal{P},\\
\mathcal{T} \Lambda(\theta, \beta)  &= \mathcal{\hat{O}}\left(\Lambda(\theta, -\beta)\right) \mathcal{T}.
\end{split}
\end{equation}
This is precisely the behavior of a semi-direct product (see appendix~(\ref{SemiDirect}) for review): commuting a first kind of group element past a second kind, results in an automorphism being applied to the first kind of group element. This observation tells us explicitly how to build the semi-direct product. We are seeking a map $\psi$ that takes elements of the discrete reflection group, and maps them to automorphisms of $\operatorname{SO}^+(1,3)$:
\begin{equation}
\psi: \operatorname{K}_4 \mapsto \operatorname{Aut}(\operatorname{SO}^+(1,3)) .
\end{equation}
Let $\psi$ be the following map:
\begin{equation}
\psi: \{1, \mathcal{P}, \mathcal{T}, \mathcal{PT}\} \mapsto \{1, \mathcal{\hat{O}}, \mathcal{\hat{O}}, 1\} ,
\end{equation}
where $\hat{\mathcal{O}}$ is the single non-trivial outer-automorphism of the Lorentz group. On a single object $V$ the semi-direct product group acts unsurprisingly (here the reflections are understood to act first, followed by the Lorentz transforms):
\begin{equation}
\left(\Lambda, \mathcal{P}\right) V = \Lambda \left(\mathcal{P}V\right).
\end{equation}
Now consider the multiplication of two elements of this semi-direct product, with reflections $\mathcal{K}, \mathcal{R} \in \operatorname{K}_4$ and Lorentz transforms $\Lambda \in \operatorname{SO}^+(1,3)$:
\begin{equation}
\begin{split}
\hspace{-1em}&\quad\>\left(\Lambda(\theta, \beta), \>\mathcal{K}\right) \cdot \left(\Lambda(\phi, \nu), \>\mathcal{R}\right)\\
\hspace{-1em}\vspace{0.5ex}&= \left(\Lambda(\theta, \beta) \cdot \psi_\mathcal{K} \left(\Lambda(\phi, \nu)\right), \mathcal{K} \cdot \mathcal{R}\right)\\
\hspace{-1em}&= \begin{cases}
\left(\Lambda(\theta, \beta) \Lambda(\phi, -\nu), \>\mathcal{K} \mathcal{R}\right) & \mathrm{if }\quad  \mathcal{K} = \mathcal{P} \> \>\mathrm{or}\> \> \mathcal{T} ,\\
\left(\Lambda(\theta, \beta) \Lambda(\phi, \nu), \>\mathcal{K} \mathcal{R}\right) & \mathrm{if }\quad  \mathcal{K} = 1 \>\> \> \mathrm{or}\>\>  \mathcal{PT}. \\
\end{cases}
\end{split}
\end{equation}
This may look busy, but all it is saying is reflections and (proper orthochronus) Lorentz transforms merely compound in the order they are applied, with the only added complication being that an odd number of reflections will invert any boost parameters of Lorentz transforms which precede it. This semi-direct product we've constructed is precisely the structure we concluded above in Eq.~(\ref{obvsemi}). It is clear the entire Lorentz group can be understood as the semi-direct product between $\operatorname{SO}^+(1,3)$ and $\operatorname{K}_4$
\begin{equation}
\operatorname{O}(1,3) = \operatorname{SO}^+(1,3)\rtimes \operatorname{K}_4,
\end{equation}
and the objects of our theory can be understood to take simultaneous representations of both these groups, and so the representation theory we have built up to this point is justified.
%\vspace{-0.8em}

\newpage
\bibliographystyle{plain} % We choose the "plain" reference style
\bibliography{Entire} % Entries are in the refs.bib file

\begin{thebibliography}{10}

\bibitem{ALBERT2003}
David Albert.
\newblock {\em Time And Change}.
\newblock Harvard University Press., 2003.

\bibitem{vier}
M.~A. Armstrong.
\newblock {\em Groups and Symmetry}.
\newblock Springer Verlag, 1988.

\bibitem{GREAVESARNTZENIUS}
Frank Arntzenius and Hilary Greaves.
\newblock Time reversal in classical electromagnetism.
\newblock {\em The British Journal for the Philosophy of Science}, 60(3), 2009.

\bibitem{Callender}
Craig Callender.
\newblock Is time ‘handed’ in a quantum world?
\newblock {\em Proceedings of the Aristotelian Society}, 100:247--269, 2000.

\bibitem{AbsAlg}
Joseph Gallian.
\newblock {\em Contemporary Abstract Algebra}.
\newblock Brooks Cole, 7 edition, 2009.

\bibitem{JACKSON3RD}
John~David Jackson.
\newblock {\em Classical Electrodynamics, Third Edition}.
\newblock John Wiley \& Sons, Inc., 1999.

\bibitem{Pontry}
A.~A. Kirillov.
\newblock {\em Elements of the Theory of Representations}.
\newblock Grundlehren der mathematischen Wissenschaften, A Series of Comprehensive Studies in Mathematics, 220. Springer Berlin Heidelberg, Berlin, Heidelberg, 1st ed. edition, 1976.

\bibitem{MALAMENT2004}
David~B. Malament.
\newblock On the time reversal invariance of classical electromagnetic theory.
\newblock {\em Studies in History and Philosophy of Science Part B: Studies in History and Philosophy of Modern Physics}, 35(2):295--315, 2004.

\bibitem{ROBERTS2022}
Bryan~W. Roberts.
\newblock {\em Reversing the Arrow of Time}.
\newblock Cambridge University Press, 2022.

\bibitem{WuKiTung}
Wu-Ki Tung.
\newblock {\em Group Theory in Physics}.
\newblock World Scientific Publishing Company, 1 edition, 1985.

\bibitem{Weinberg_1972}
Steven Weinberg.
\newblock {\em Gravitation and Cosmology}.
\newblock John Wiley \& sons, Inc, 1972.

\bibitem{Weinberg_1995}
Steven Weinberg.
\newblock {\em The Quantum Theory of Fields}.
\newblock Cambridge University Press, 1995.

\bibitem{Wigner1932}
E.~Wigner.
\newblock Ueber die operation der zeitumkehr in der quantenmechanik.
\newblock {\em Nachrichten von der Gesellschaft der Wissenschaften zu Göttingen, Mathematisch-Physikalische Klasse}, 1932:546--559, 1932.

\bibitem{WIGNERBOOK}
Eugine~P. Wigner.
\newblock {\em Group Theory and its Application to the Quantum Mechanics of Atomic Spectra}.
\newblock Academic Press, 1959.

\end{thebibliography}

%\end{changemargin}
\end{document}